\documentclass[aps,prl,reprint,superscriptaddress]{revtex4-1}
\usepackage{lineno,xcolor}
\usepackage{graphicx}
\usepackage{multirow}
\usepackage{float}
\usepackage{bm}
\usepackage[breaklinks=true]{hyperref}
\usepackage{enumitem}
\usepackage{epstopdf}
\usepackage{amsmath}
\usepackage{amsfonts}
\usepackage{verbatim}   
\usepackage{mathtools} 
\usepackage{bm} 
\usepackage{lmodern}
\usepackage[english]{babel}
\usepackage{xcolor}      
\usepackage{siunitx}     
\usepackage{gensymb} 
\usepackage{tikz}
\usepackage{soul}
\usepackage{amssymb}%

\providecommand{\U}[1]{\protect\rule{.1in}{.1in}}
\hypersetup{bookmarksnumbered, pdfpagemode=UseOutlines,
pdfauthor={A.\ Aqeel},
pdftitle={Electrical detection of skyrmions in a chiral magnetic insulator}, pdfdisplaydoctitle,
colorlinks=true, citecolor=blue, filecolor=blue, linkcolor=blue, urlcolor=blue}

\usepackage{hyperref}
\usepackage[bottom]{footmisc}

\begin{document}
\date{\today}
\title{All-electrical detection of skyrmion lattice state and chiral surface twists}
\author{A.\ \surname{Aqeel}}
\affiliation{Department of Physics, Technical University Munich, 85748 Garching, Germany}
\author{M.\ \surname{Azhar}}
\affiliation{Zernike Institute for Advanced Materials, University of Groningen, Nijenborgh 4, 9747 AG Groningen, The Netherlands}
\author{N.\ \surname{Vlietstra}}
\affiliation{Department of Physics, Technical University Munich, 85748 Garching, Germany}
\affiliation{Walther-Mei$\ss$ner-Institute, 85748 Garching, Germany}
\author{A.\ \surname{Pozzi}}
\affiliation{Zernike Institute for Advanced Materials, University of Groningen, Nijenborgh 4, 9747 AG Groningen, The Netherlands}
\author{J.\ \surname{Sahliger}}
\affiliation{Department of Physics, Technical University Munich, 85748 Garching, Germany}
\author{H.\ \surname{Huebl}}
\affiliation{Walther-Mei$\ss$ner-Institute, 85748 Garching, Germany}
\affiliation{Department of Physics, Technical University Munich, 85748 Garching, Germany}
\affiliation{Munich Center for Quantum Science and Technology (MCQST), Schellingstr. 4, D-80799 Munich,Germany}
\author{T.\ T.\ M.\ \surname{Palstra}}
\affiliation{Zernike Institute for Advanced Materials, University of Groningen, Nijenborgh 4, 9747 AG Groningen, The Netherlands}
\author{C.\ H.\ \surname{Back}}
\affiliation{Department of Physics, Technical University Munich, 85748 Garching, Germany}
\affiliation{Munich Center for Quantum Science and Technology (MCQST), Schellingstr. 4, D-80799 Munich,Germany}
\author{M.\ \surname{Mostovoy}}
\affiliation{Zernike Institute for Advanced Materials, University of Groningen, Nijenborgh 4, 9747 AG Groningen, The Netherlands}

\begin{abstract}
We study the high-temperature phase diagram of the chiral magnetic insulator Cu$_2$OSeO$_3$ by measuring the spin-Hall magnetoresistance (SMR) in a thin Pt electrode. 
We find distinct changes in the phase and amplitude of the SMR signal at critical lines separating different  magnetic phases of bulk Cu$_2$OSeO$_3$. 
The skyrmion lattice state appears as a strong dip in the SMR phase.
A strong enhancement of the SMR amplitude is observed in the conical spiral state, which we explain by an additional symmetry-allowed contribution to the SMR present in non-collinear magnets.
We demonstrate that the SMR can be used as an all-electrical probe of chiral surface twists and skyrmions in magnetic insulators. 
\end{abstract}
\keywords{}\maketitle

\renewcommand\linenumberfont{\normalfont\tiny\sffamily\color{gray}}
Magnetic skyrmions are nano-scale spin-swirling objects of great interest due to their small size and topological protection~\cite{Nagaosa2013,Fert2017}. 
These magnetic solitons are being considered as promising candidates for information bits in ultra compact thin film memory devices where surfaces and interfaces play a crucial role. 
Recent theoretical studies of chiral magnets showed that their surface magnetization can be strongly modified by the bulk Dzyaloshinskii-Moriya interactions (DMI), leading to a surface structure which deviates from the bulk skyrmion lattice. These so-called chiral surface twists are expected to be prominent in the field-polarized magnetic state restraining the magnetization to fully saturate near the surface along the applied field direction~\cite{Rybakov2013}. They also affect helicity and stability of three-dimensional skyrmions~\cite{Rybakov2013,Meynell2014,Rybakov2015,Leonov2016PRL} and first experimental evidence for the existence of these chiral surface twists has been recently reported for skyrmion tubes aligned perpendicular to the sample surface~\cite{Zhang2018}. Chiral twists were also observed in thin film multilayers~\cite{Legrand_2018,Zhang2020} potentially relevant for skyrmionics applications.

In this paper, we report all-electrical detection of skyrmions in Cu$_2$OSeO$_3$ (CSO). Close to spin-ordering temperature, this chiral magnetic insulator with P$2_13$ symmetry hosts crystals of three-dimensional tubes formed by stacked Bloch-type skyrmions~\cite{Seki2012}. The effect of the chiral twist on the electrical fingerprint of the skyrmion tubes, which in our device are parallel to the CSO surface, has not been studied so far. 

The detection of these surface modifications is difficult, yet important for integration of skyrmions in multi-layered devices. Here, we systematically track the surface magnetization by a detailed investigation of the angular dependence of the spin Hall magnetoresistance (SMR) in the high-temperature part \((40~{\rm K} - 60~{\rm K})\) of the CSO phase diagram.
\begin{figure}[!t]
    \includegraphics[width = 0.5\textwidth,keepaspectratio]{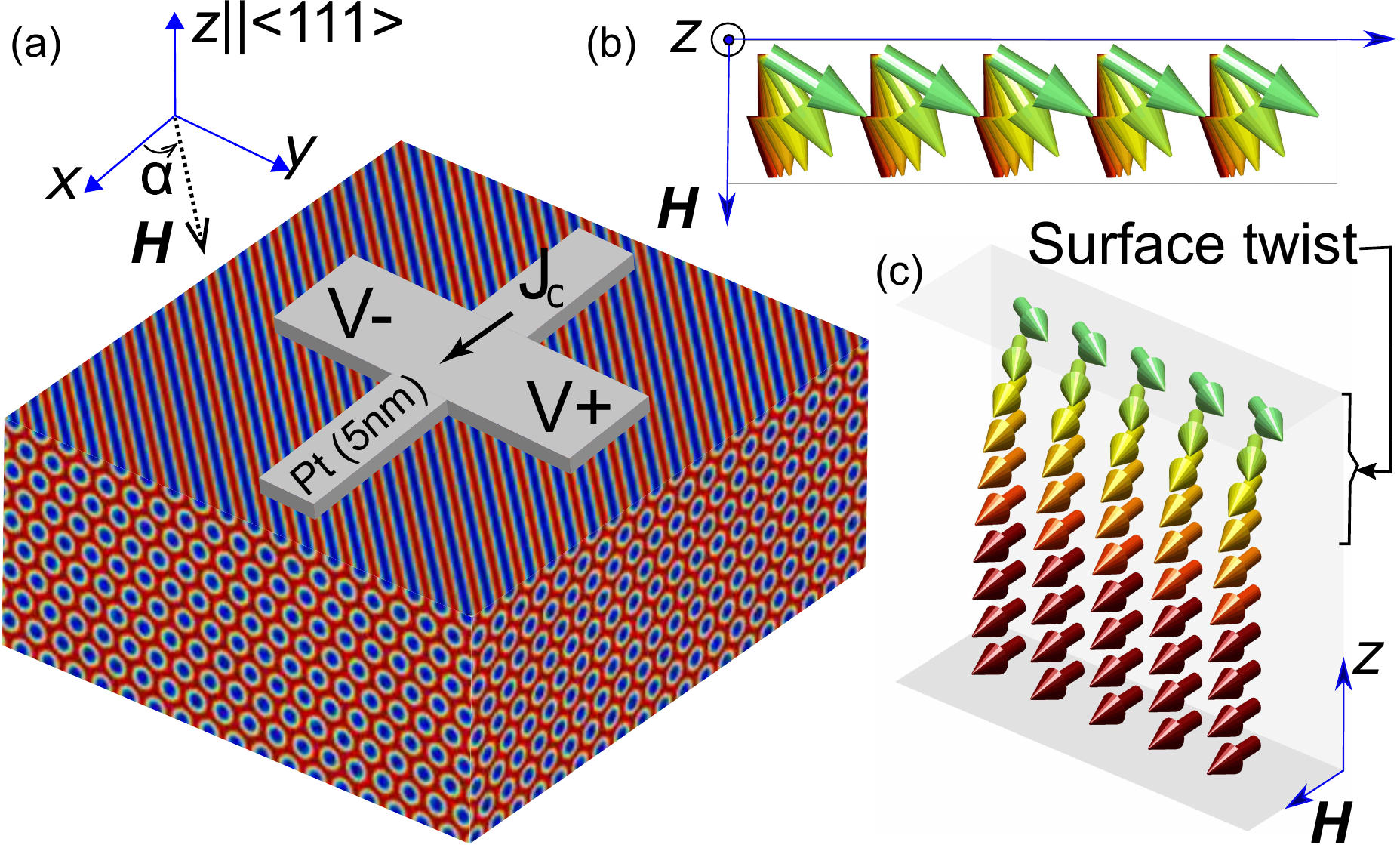}

\caption{{\bf (a)} Schematic representation of the device configuration used to measure the spin-Hall magnetoresistance in the Pt$|$CSO bilayer system, with an applied magnetic field $H$ at an angle $\alpha$ to $x$, the direction of the current. The skyrmion tubes are represented by color-coding the magnetization projection along the applied magnetic field direction (red to blue). Top {\bf (b)} and cross-sectional {\bf (c)} view of the surface twists in the field-polarized state. The increase in the twist angle between the magnetization and the applied field near the surface is indicated by color (red to green). 
}
\label{fig:1}%
\end{figure}
SMR has been used in the past to electrically detect the in-plane magnetization of collinear magnetic insulators~\cite{Vlietstra2013,Nakayama2013,Althammer2013}, non-collinear magnets~\cite{Aqeel2015,Ganzhorn2016,Aqeel2016,Aqeel2017}, and antifermagnetic materials~\cite{Ji2017,Hoogeboom2017,Wang2017,Fischer2018,Lebrun2019}. In the SMR, the resistance of a heavy metal (Pt) is sensitive to the magnetization direction of an adjacent magnetic layer. In collinear magnets, the SMR is theoretically approximated by the magnetic moment density at the interface~\cite{JiaSTT2011,Chen2013}. In addition to a longitudinal resistance change, a transverse voltage arises given by 
\begin{equation}
V_{\rm SMR} \propto m_{x}m_{y},
\label{eq:1}
\end{equation} 
 where $m_x$ and $m_y$ are the in-plane components of the unit vector $\bm m$ describing the magnetization direction. For non-collinear magnets with a spin relaxation length, $\xi$, much smaller than the typical length scale of variation of $\bm m(\bm x)$, the right-hand side of Eq.\eqref{eq:1} is replaced by its average over the interface ~\cite{Aqeel2016}:

\begin{equation}
V_{\rm SMR} \propto \langle m_{x}m_{y} \rangle .
\label{eq:2}
\end{equation} 
When the external magnetic field $H$ (larger than the saturation magnetic field $H_{\rm c2}$) is rotated in the plane of the normal metal$|$magnet interface, the SMR voltage, $V_{\rm SMR}$, follows a sinusoidal angular dependence with a periodicity of 180$^\circ$ (see Fig.~\ref{fig:2}(a)). This behavior of the SMR is expected when the magnetization is fully aligned with the applied magnetic field. However, any in-plane tilt of the magnetization away from $H$ will give rise to an additional phase, $\phi$, in the angular dependence of $V_{\rm SMR}$ making this physical parameter very sensitive to the magnetic structure at the interface.

\begin{figure}[!htb]
 \includegraphics[width = 0.41\textwidth,keepaspectratio]{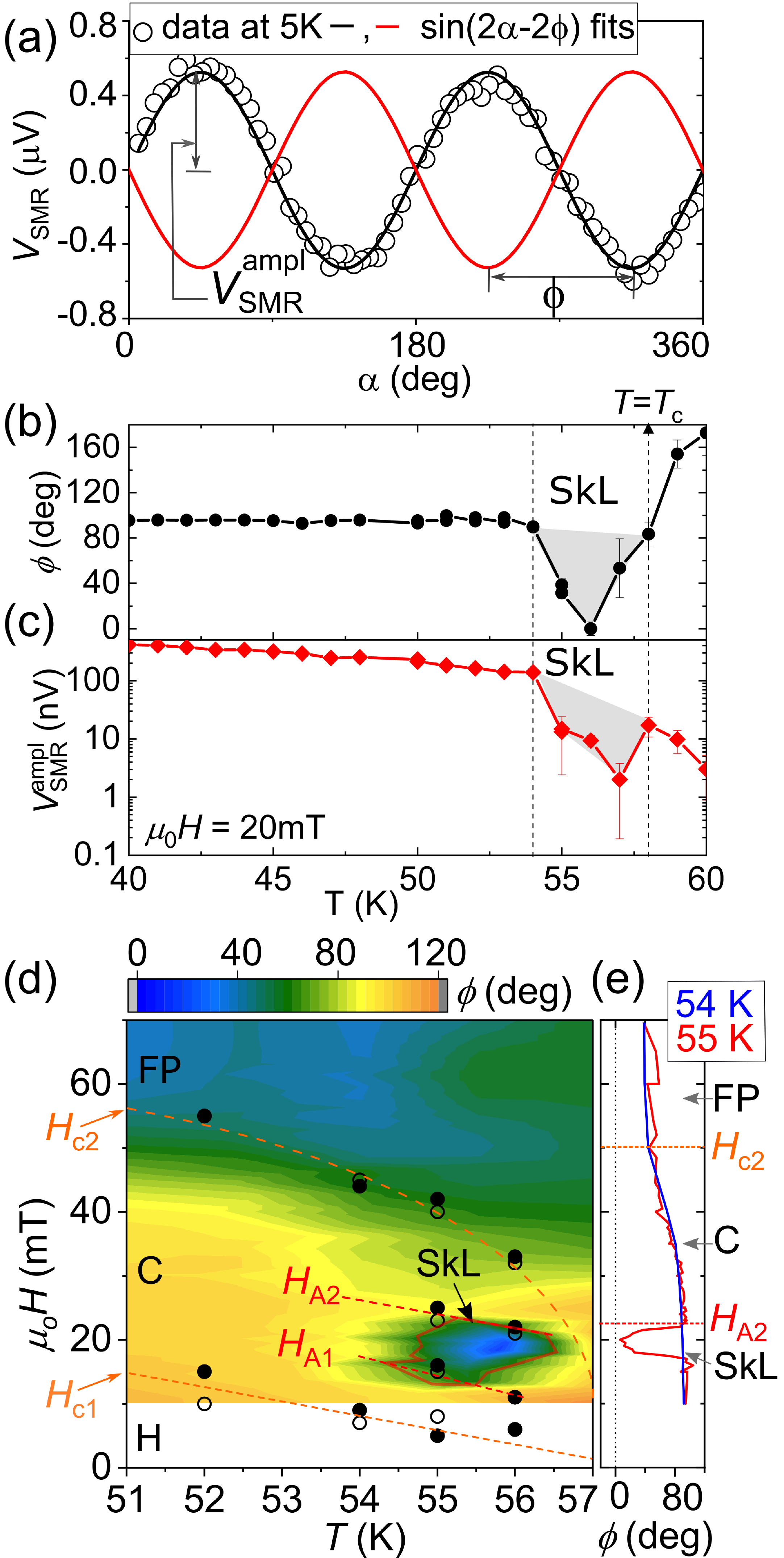}
\caption{{\bf (a)} Angular dependence of the SMR signal, $V_{\rm SMR}$, measured at $\mu_0H = 600$~mT in a planar-Hall geometry shown in Fig.~\ref{fig:1}(a). $V_{\rm SMR}^{\rm ampl}$ and $\phi$ are the amplitude and phase of $V_{\rm SMR}$, respectively. The solid lines show the $V_{\rm SMR}^{\rm ampl} \sin$(2$(\alpha - \phi)$) fit with $\phi=0^\circ$ (black) and $\phi=90^\circ$ (red). {\bf (b)} and {\bf (c)} show the temperature dependence of $V_{\rm SMR}^{\rm ampl}$ and $\phi$, respectively. Here, the shaded area corresponds to the skyrmion lattice state. {\bf (d)} shows the contour map of phase $\phi$ versus magnetic field and temperature combined with the results of broadband ferromagnetic resonance. The hollow and solid circles represent the phase boundaries extracted from the magnetic resonance data measured by applying $\bm H$ along the [1 -1 1] and [-2 -1 1] crystallographic directions of CSO, respectively. The magnetic phase boundaries are highlighted by dashed lines. We identify helical (H), conical (C) ($H_{\rm c1}$ is the low field boundary), field-polarized ferrimagnetic state (FP) ($H_{\rm c2}$ is the low field boundary) and the skyrmion lattice state (SkL) ($H_{\rm A1}$ and $H_{\rm A2}$ are low and high field boundaries). {\bf (e)} shows the line scans of $\phi$ as function of applied field $H$ at two different temperatures.
}
\label{fig:2}%
\end{figure}

 It is worth noting that previous SMR studies of CSO were performed outside the stability region of the Skyrmion lattice (SkL) phase~\cite{Aqeel2016,Aqeel2017}. The measurements reported here focus on the SkL. We show that the SMR sensitively measures the deviation of the local interface magnetization from $\bm H$, allowing us to probe the highly non-collinear SkL spin texture electrically.
 The obtained results provide new insights into electrical detection of skyrmions and their surface deformations in magnetic insulators.

SMR measurements are performed in the device configuration shown in Fig.~\ref{fig:1}(a), where a 5~nm thick Pt Hall cross is structured onto an oriented polished CSO cuboid (dimensions: 4$\times$4$\times$2~mm$^3$). A low frequency ac-current (\(f=17\)~Hz) is applied through Pt along the x axis. The SMR is detected in the transverse configuration (along the y axis) by measuring the first harmonic voltage response $V_1$ using a lock-in amplifier~\cite{Vlietstra2014} (see Supplementary Material for more details). The angular dependence of the SMR is recorded by rotating the applied magnetic field within the xy-plane parallel to the Pt/CSO interface. Here, the assumption that the SKL follows the field direction, has been recently confirmed in CSO for an in-plane field rotation in a different crystal orientation~\cite{Zhang2020}. A typical example of the angular dependence of the SMR data measured at 5 K in the ferrimagnetic state of CSO ($\mu_0H = 600$~mT) is shown in Fig.~\ref{fig:2}(a). The amplitude $V_{\rm SMR}^{\rm ampl}$ and phase $\phi$ are extracted by fitting the data with \(V_{\rm SMR}= V_{\rm SMR}^{\rm ampl} \sin(2(\alpha-\phi))\). Here, $\alpha$ is the angle between the applied current $I$ and the applied magnetic field, $\bm H$, as defined in Fig.~\ref{fig:1}(a). 
The phase $\phi$ is zero, if magnetization is parallel to the field. The $\sin(2\alpha)$ dependence with zero phase has been observed in the ferrimagnetic state of CSO at 5K (see Fig.~\ref{fig:2}(a)). On the other hand, in the helical and low-field conical states, the magnetic moments align almost perpendicular to the propagation vector of the spiral, resulting in $\phi\sim 90^\circ$~\cite{Aqeel2016}.

To identify the skyrmion pocket, we recorded the angular dependence of the SMR using two sets of measurement protocols: (i) at fixed fields as a function of temperature (T-scan) and (ii) at fixed temperature, recording the SMR angular dependence at various magnetic fields strengths (H-scan). Figures~\ref{fig:2}(b) and \ref{fig:2}(c) show the T-scan of $\phi$ and $V_{\rm SMR}^{\rm ampl}$ extracted from the angular dependence of the SMR signal as exemplified in Fig.~\ref{fig:2}(a). At $\mu_0H = 20$~mT outside the skyrmion pocket, CSO is expected to be in the conical spiral state, in which $\phi \sim 90^\circ$ is expected. Indeed, we observe that in a wide temperature range $\phi \sim 90^\circ$ except near $T_{\rm c}$ where a clear dip in the phase $\phi$ is found (labeled as SkL in Fig.~\ref{fig:2}(b)). A similar signature dip is observed in $V_{\rm SMR}^{\rm ampl}$ (see Fig.~\ref{fig:2}(c); note the logarithmic voltage scale). 

	\begin{figure}[h]
	\centering
	\includegraphics[width = 0.5\textwidth,keepaspectratio]{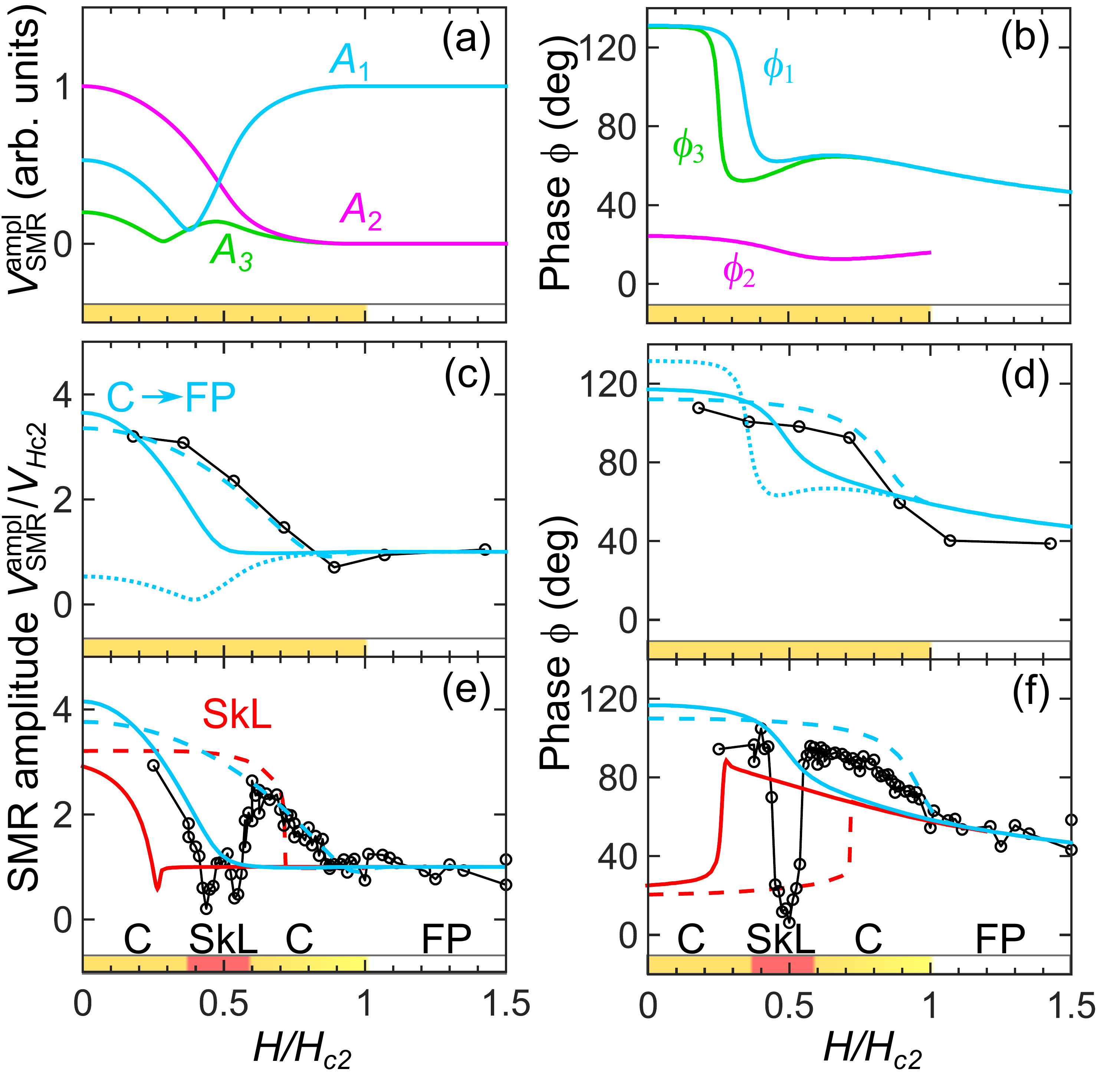}
	\caption{Field dependence of the spin-Hall magnetoresistance {\bf (a)} amplitude, $V_{\rm SMR}^{\rm ampl}$, and {\bf (b)} phase, $\phi$, in the conical spiral (C) and collinear field-polarized (FP) states calculated for each of the three terms in Eq.~\eqref{eq:RSMR3} separately: blue ($A_1$ and $\phi_1$, due to the 1$^{\rm{st}}$ term), magenta ($A_2$ and $\phi_2$, due to the 2$^{\rm{nd}}$ term) and green lines ($A_3$ and $\phi_3$, due to the 3$^{\rm{rd}}$ term). Field dependence of {\bf (c)} $V_{\rm SMR}^{\rm ampl}$ and {\bf (d)} $\phi$ in the C and FP states calculated numerically using Eq.~\eqref{eq:RSMR3} with $bQ/a=-1.59$ and $c/a=0$ (solid blue lines). The dashed line shows the effect of the interfacial DMI, for $D_{\rm int}=-0.30 J$. The dotted blue line is obtained by considering only the first term in Eq.~\eqref{eq:RSMR3}. Open circles are the experimental data taken at 51 K. Here, the critical field, $H_{\rm c2}$ = 56 mT. Field-dependence of {\bf (e)} $V_{\rm SMR}^{\rm ampl}$ and {\bf (f)} $\phi$ calculated numerically using Eq.~\eqref{eq:RSMR3} with $bQ/a=-1.84$ and $c/a=0$ for the C and FP states (blue solid lines) and for the skyrmion lattice (SkL) state (red solid lines). The dashed blue (red) lines show the effect of the interfacial DMI of strength $D_{\rm int}=-0.40 J$ for the C (SkL) state. Open circles are results of experimental measurements at 55 K. The color-bars at the $x$ axes of the figures indicate the magnetic phases, deduced from FMR measurements for the experimental data: the C, SkL and FP state is indicated by yellow, red and white color, respectively. The critical fields determined from FMR measurements at 55 K are: $H_{\rm c2}=40$ mT, and the boundaries of the SkL, $H_{\rm A1}$ and $H_{\rm A2}$, are 15 mT and 23 mT, respectively.}
	\label{fig:3}
	\end {figure}

To outline the boundaries of the skyrmion pocket, we constructed the high-temperature phase diagram of CSO by mapping the phase $\phi$ as a function of field and temperature using H-scans. Figure~\ref{fig:2}(d) shows the obtained phase diagram along with the results of ferromagnetic resonance (FMR) measurements which are used here to identify the phase boundaries of the different magnetic states of the CSO crystal. To measure the FMR response, a broad band spin-wave spectroscopy technique~\cite{Schwarze_2015} is used~(see the Supplementary Material for details). The phase boundaries of the FMR data (shown by circles in Fig.~\ref{fig:2}(d)) agree well with those extracted from the SMR phase $\phi$. $V_{\rm SMR}^{\rm ampl}$ also exhibits distinct anomalies at the magnetic phase transitions, though less pronounced than those in the SMR phase $\phi$ (see 
Fig.~S3 in the Supplementary Material). From the experimentally obtained phase diagram shown in Fig.~\ref{fig:2}(d), the main observations are as follows: (i) the SkL state is clearly distinguishable from other magnetic states of CSO, and (ii) close to $T_{\rm c}$, the phase $\phi$ remains non-zero above $H_{c2}$. To comprehend these findings, two H-scans of $\phi$ at two temperatures are shown in Fig.~\ref{fig:2}(e). At 54~K, the phase $\phi$ is almost 90$^\circ$ in the helical state and decreases in the conical state of CSO; the lowest values of the phase $\phi$ are observed in the FM state. At 55~K, the phase $\phi$ follows the same trend as observed at 54~K and in addition shows a significant change in the SkL state of CSO. Importantly, in the collinear state ($H \ge H_{\rm c2}$), $\phi$ remains non-zero ($\sim 50^\circ$ at $\mu_0H$ = 50 mT). 

A nonzero SMR phase above $H_{\rm c2}$ provides direct evidence for a chiral surface twist, i.e. 
the tilt of the magnetization at the interface away from the magnetic field direction \cite{Rybakov2013,Meynell2014,Rybakov2015,Leonov2016PRL}. The observed field-dependence of $\phi$ can be understood using Eq.~\eqref{eq:2} and taking into account the modification of the conical spiral near the surface. We numerically calculated this modification by minimizing the energy of a classical spin model describing CSO with periodic boundary conditions along the in-plane directions and open boundary conditions along the surface normal direction (dotted blue line in Fig.~\ref{fig:3}(d); see Supplementary Material for details of the calculations).
This approach reproduces the experimentally observed decrease of the SMR phase $\phi$ with increasing magnetic field in the conical spiral state and a nonzero value of $\phi$ at the transition to the field-induced collinear state: due to the rotation of spins at the interface away from the magnetic field direction, $\phi = \ang{60}$ at $H=H_{\rm c2}$ and then slowly decreases with increasing magnetic field for $H>H_{\rm c2}$, in agreement with our observations (cf. measured data (open circles) with dotted blue line in Fig.~\ref{fig:3}(d)). 

However, this approach fails to explain the observed field-dependence of the SMR amplitude. 
The calculated $V_{\rm SMR}^{\rm ampl}$ reaches its maximal value at $H_{\rm c2}$ and stays constant at higher fields (dotted blue line in Fig.~\ref{fig:3}(c)). In contrast, the SMR amplitude measured above 50~K is relatively small at $H \ge H_{\rm c2}$. It reaches a maximum at rather low magnetic fields and then decreases with increasing field (see Fig.~\ref{fig:3}(c)). The origin of this discrepancy can be traced back to Eq.~\eqref{eq:2} that only involves in-plane magnetization components which become on average larger as the strength of the magnetic field parallel to the interface increases. As a result, the calculated SMR amplitude grows, as $H$ tends towards $H_{\rm c2}$. This discussion shows that the surface twist alone cannot explain the high-temperature SMR data and motivates us to add new phenomenological terms to Eq.~\eqref{eq:2} allowed by symmetry of the polycrystalline Pt film (rotations around the normal to the film and vertical mirrors):
\begin{equation}
\begin{split}
\rho_{\rm SMR}^\perp = \quad & a <m_{x}m_{y}>+b<m_{z}\overset{\leftrightarrow}{\partial_y}m_{x}> \\ \\
& +c<m_{z}^2m_{x}m_{y}>,
\end{split}
\label{eq:RSMR3}
\end{equation}
where $\rho_{\rm SMR}^\perp$ is the transverse SMR resistivity. The first term coincides with Eq.~\eqref{eq:2}. The second term is the contribution proportional to the magnetization gradient along the in-plane direction: 
$m_{z}\overset{\leftrightarrow}{\partial_y}m_{x} = m_{z} (\partial_y m_{x}) - (\partial_y m_{z}) m_{x}$. This term is sensitive to the presence of the helical spin spiral modulation at the interface, and is proportional to the spiral wavevector, $Q=2\pi/\lambda$, where $\lambda$ is the period of the spiral. The last term proportional to the fourth power of magnetization can originate from higher-order spin torques \cite{Hanke2020} (the higher-order terms in $\bm m$ are omitted, for simplicity). The microscopic mechanism behind the second term in Eq.~(\ref{eq:RSMR3}) is unclear. According to SMR theory~\cite{Chen2013}, it should be a $\xi Q$ correction to the first term, $\xi \sim 1$ nm being the relaxation length in Pt, whereas our experiment suggests that the contributions of first and second terms of Eq.~(\ref{eq:RSMR3}) are comparable. 

The field-dependence of these three contributions to the SMR amplitude and phase is shown in Figs.~\ref{fig:3}(a,b). 
Both the second and the third terms can explain the observed decrease of $V_{\rm SMR}^{\rm ampl}$ with increasing magnetic field. However, the third term gives rise to an extra minimum, which is not observed in experiment. The best fit of the field-dependence of both the amplitude and phase of SMR is obtained for \( bQ/a =-1.59 \), \(c = 0\) at 51 K and \(bQ/a=-1.84\), \(c = 0\) at 55 K (solid blue lines in Fig.~\ref{fig:3}(c-f)). 
The surprising conclusion is that the contribution of the new term proportional to the magnetization gradient (the second term in Eq.~\eqref{eq:RSMR3}) to the SMR of CSO is comparable to that of the first term in Eq.~\eqref{eq:RSMR3} that works well for collinear magnets.

Although the additional SMR term improves the overall agreement between the calculated and experimental SMR curves, the conical spiral state in the theoretical plots seems to disappear at a field that is lower than $H_{\rm c2}$ deduced from our SMR and FMR measurements (see solid blue lines in Fig.~\ref{fig:3}(c-f)). This mismatch reflects an interesting effect found by numerical simulations: in addition to the rotation of spins around the surface normal, present in both the collinear and conical states, the conical angle at the interface is smaller than that in bulk. Moreover, it becomes very small at fields well below $H_{\rm c2}$, so that, in practice, the interface has a lower critical field than the bulk 
(see Figs. S6 and S7
in the Supplementary Material).

Our simulations also reveal strong sensitivity of the SMR to additional spin interactions at the interface, such as the interfacial Dzyaloshinskii-Moriya interaction (iDMI) enabled by inversion symmetry breaking at Pt$|$CSO interface and the strong spin-orbit coupling of Pt. The iDMI with a positive (negative) constant, $D_{\rm int}$, decreases (increases) the cone angle of the spiral at the interface, leaving the surface twist angle practically unchanged. Hence, it affects the effective critical field at the interface, which can solve the above mentioned problem. Figures~\ref{fig:3}(c-f) show that iDMI can improve the agreement between the theoretically calculated SMR amplitude and phase (blue dashed line) and the experimental data. The calculations are done for $D_{\rm int}=-0.40J$ at 55 K and $D_{\rm int}=-0.30J$ at 51 K, where $J$ is the nearest-neighbor Heisenberg exchange constant. 

Finally, we discuss the sudden changes in the experimentally measured amplitude and phase of the SMR associated with the intervening skyrmion lattice phase. Figures~\ref{fig:3}(e,f) show the field dependence of $V_{\rm SMR}^{\rm ampl}$ and $\phi$ in the SkL state calculated numerically using Eq.~\eqref{eq:RSMR3} with and without the iDMI (dashed and solid red lines, respectively). 
Although minimization of the energy of stable and metastable spin configurations at zero temperature does not allow us to obtain boundaries of the SkL phase near $T_c$, our calculations show that the SMR in the SkL state can be very different from that in the conical spiral state. In particular, the calculated phase $\phi$ in the SkL state is smaller than that in the conical spiral state at low magnetic fields, which agrees with the observed dip in the SMR phase. In addition, we find a sudden jump in both the phase and the amplitude of the SMR in the SkL state related to an abrupt change of the skyrmion crystal spin configuration at the surface (See 
Fig.~S9 in Supplementary Materials). At high fields, skyrmions are repelled from the edge and spins at the interface are collinear and nearly parallel to the surface. At low fields, the skyrmion centers are residing at the interface, which makes their topological charge smaller than 1 \cite{Keesman_2015}. The field at which this change occurs is strongly affected by the interfacial DMI (red dashed lines in Fig.~\ref{fig:3}).

To summarize, our SMR measurements provide a clear evidence for a surface twist in the conical and collinear phases near $T_c$ and make it possible to \textit{all-electrically} outline the boundaries of the skyrmion crystal phase, at which both the amplitude and the phase of the SMR show a profound discontinuity. The theoretical description of SMR for non-collinear chiral magnets, such as CSO, is more involved than that for collinear ferromagnets. First, SMR is affected by subtle interfacial effects, such as the difference between the surface and bulk critical fields. Second, a new phenomenological term in the SMR expression was required to reproduce the field-dependence of the SMR amplitude. In addition, the SMR was found to be sensitive to interfacial interactions. Our experimental and theoretical results conclusively show that the SMR, as a probe of the surface magnetization, is an important tool for the all-electrical detection of magnetic phase transitions in chiral magnets.


\begin{acknowledgments}
We would like to thank B. J. van Wees, T. Kuschel and J. Bass for useful discussions. 
This work has been funded by the Deutsche Forschungsgemeinschaft (DFG, German Research Foundation) via project number 107745057 - TRR 80, via SPP2137, as well as via Germany’s Excellence Strategy – EXC-2111 – 390814868, Vrije FOM-programma `Skyrmionics' and the Zernike Institute for Advanced Materials. 
\end{acknowledgments}




\begin{thebibliography}{28}%
\makeatletter
\providecommand \@ifxundefined [1]{%
 \@ifx{#1\undefined}
}%
\providecommand \@ifnum [1]{%
 \ifnum #1\expandafter \@firstoftwo
 \else \expandafter \@secondoftwo
 \fi
}%
\providecommand \@ifx [1]{%
 \ifx #1\expandafter \@firstoftwo
 \else \expandafter \@secondoftwo
 \fi
}%
\providecommand \natexlab [1]{#1}%
\providecommand \enquote  [1]{``#1''}%
\providecommand \bibnamefont  [1]{#1}%
\providecommand \bibfnamefont [1]{#1}%
\providecommand \citenamefont [1]{#1}%
\providecommand \href@noop [0]{\@secondoftwo}%
\providecommand \href [0]{\begingroup \@sanitize@url \@href}%
\providecommand \@href[1]{\@@startlink{#1}\@@href}%
\providecommand \@@href[1]{\endgroup#1\@@endlink}%
\providecommand \@sanitize@url [0]{\catcode `\\12\catcode `\$12\catcode
  `\&12\catcode `\#12\catcode `\^12\catcode `\_12\catcode `\%12\relax}%
\providecommand \@@startlink[1]{}%
\providecommand \@@endlink[0]{}%
\providecommand \url  [0]{\begingroup\@sanitize@url \@url }%
\providecommand \@url [1]{\endgroup\@href {#1}{\urlprefix }}%
\providecommand \urlprefix  [0]{URL }%
\providecommand \Eprint [0]{\href }%
\providecommand \doibase [0]{http://dx.doi.org/}%
\providecommand \selectlanguage [0]{\@gobble}%
\providecommand \bibinfo  [0]{\@secondoftwo}%
\providecommand \bibfield  [0]{\@secondoftwo}%
\providecommand \translation [1]{[#1]}%
\providecommand \BibitemOpen [0]{}%
\providecommand \bibitemStop [0]{}%
\providecommand \bibitemNoStop [0]{.\EOS\space}%
\providecommand \EOS [0]{\spacefactor3000\relax}%
\providecommand \BibitemShut  [1]{\csname bibitem#1\endcsname}%
\let\auto@bib@innerbib\@empty
\bibitem [{\citenamefont {Nagaosa}\ and\ \citenamefont
  {Tokura}(2013)}]{Nagaosa2013}%
  \BibitemOpen
  \bibfield  {author} {\bibinfo {author} {\bibfnamefont {N.}~\bibnamefont
  {Nagaosa}}\ and\ \bibinfo {author} {\bibfnamefont {Y.}~\bibnamefont
  {Tokura}},\ }\href {\doibase 10.1038/nnano.2013.243} {\bibfield  {journal}
  {\bibinfo  {journal} {Nature Nanotechnology}\ }\textbf {\bibinfo {volume}
  {8}},\ \bibinfo {pages} {899} (\bibinfo {year} {2013})}\BibitemShut {NoStop}%
\bibitem [{\citenamefont {Fert}\ \emph {et~al.}(2017)\citenamefont {Fert},
  \citenamefont {Reyren},\ and\ \citenamefont {Cros}}]{Fert2017}%
  \BibitemOpen
  \bibfield  {author} {\bibinfo {author} {\bibfnamefont {A.}~\bibnamefont
  {Fert}}, \bibinfo {author} {\bibfnamefont {N.}~\bibnamefont {Reyren}}, \ and\
  \bibinfo {author} {\bibfnamefont {V.}~\bibnamefont {Cros}},\ }\href {\doibase
  10.1038/natrevmats.2017.31} {\bibfield  {journal} {\bibinfo  {journal}
  {Nature Reviews Materials}\ }\textbf {\bibinfo {volume} {2}},\ \bibinfo
  {pages} {17031} (\bibinfo {year} {2017})}\BibitemShut {NoStop}%
\bibitem [{\citenamefont {Rybakov}\ \emph {et~al.}(2013)\citenamefont
  {Rybakov}, \citenamefont {Borisov},\ and\ \citenamefont
  {Bogdanov}}]{Rybakov2013}%
  \BibitemOpen
  \bibfield  {author} {\bibinfo {author} {\bibfnamefont {F.~N.}\ \bibnamefont
  {Rybakov}}, \bibinfo {author} {\bibfnamefont {A.~B.}\ \bibnamefont
  {Borisov}}, \ and\ \bibinfo {author} {\bibfnamefont {A.~N.}\ \bibnamefont
  {Bogdanov}},\ }\href {\doibase 10.1103/PhysRevB.87.094424} {\bibfield
  {journal} {\bibinfo  {journal} {Phys. Rev. B}\ }\textbf {\bibinfo {volume}
  {87}},\ \bibinfo {pages} {094424} (\bibinfo {year} {2013})}\BibitemShut
  {NoStop}%
\bibitem [{\citenamefont {Meynell}\ \emph {et~al.}(2014)\citenamefont
  {Meynell}, \citenamefont {Wilson}, \citenamefont {Fritzsche}, \citenamefont
  {Bogdanov},\ and\ \citenamefont {Monchesky}}]{Meynell2014}%
  \BibitemOpen
  \bibfield  {author} {\bibinfo {author} {\bibfnamefont {S.~A.}\ \bibnamefont
  {Meynell}}, \bibinfo {author} {\bibfnamefont {M.~N.}\ \bibnamefont {Wilson}},
  \bibinfo {author} {\bibfnamefont {H.}~\bibnamefont {Fritzsche}}, \bibinfo
  {author} {\bibfnamefont {A.~N.}\ \bibnamefont {Bogdanov}}, \ and\ \bibinfo
  {author} {\bibfnamefont {T.~L.}\ \bibnamefont {Monchesky}},\ }\href {\doibase
  10.1103/PhysRevB.90.014406} {\bibfield  {journal} {\bibinfo  {journal} {Phys.
  Rev. B}\ }\textbf {\bibinfo {volume} {90}},\ \bibinfo {pages} {014406}
  (\bibinfo {year} {2014})}\BibitemShut {NoStop}%
\bibitem [{\citenamefont {Rybakov}\ \emph {et~al.}(2015)\citenamefont
  {Rybakov}, \citenamefont {Borisov}, \citenamefont {Bl\"ugel},\ and\
  \citenamefont {Kiselev}}]{Rybakov2015}%
  \BibitemOpen
  \bibfield  {author} {\bibinfo {author} {\bibfnamefont {F.~N.}\ \bibnamefont
  {Rybakov}}, \bibinfo {author} {\bibfnamefont {A.~B.}\ \bibnamefont
  {Borisov}}, \bibinfo {author} {\bibfnamefont {S.}~\bibnamefont {Bl\"ugel}}, \
  and\ \bibinfo {author} {\bibfnamefont {N.~S.}\ \bibnamefont {Kiselev}},\
  }\href {\doibase 10.1103/PhysRevLett.115.117201} {\bibfield  {journal}
  {\bibinfo  {journal} {Phys. Rev. Lett.}\ }\textbf {\bibinfo {volume} {115}},\
  \bibinfo {pages} {117201} (\bibinfo {year} {2015})}\BibitemShut {NoStop}%
\bibitem [{\citenamefont {Leonov}\ \emph {et~al.}(2016)\citenamefont {Leonov}
  \emph {et~al.}}]{Leonov2016PRL}%
  \BibitemOpen
  \bibfield  {author} {\bibinfo {author} {\bibfnamefont {A.~O.}\ \bibnamefont
  {Leonov}} \emph {et~al.},\ }\href {\doibase 10.1103/PhysRevLett.117.087202}
  {\bibfield  {journal} {\bibinfo  {journal} {Phys. Rev. Lett.}\ }\textbf
  {\bibinfo {volume} {117}},\ \bibinfo {pages} {087202} (\bibinfo {year}
  {2016})}\BibitemShut {NoStop}%
\bibitem [{\citenamefont {Zhang}\ \emph {et~al.}(2018)\citenamefont {Zhang},
  \citenamefont {van~der Laan}, \citenamefont {Wang}, \citenamefont
  {Haghighirad},\ and\ \citenamefont {Hesjedal}}]{Zhang2018}%
  \BibitemOpen
  \bibfield  {author} {\bibinfo {author} {\bibfnamefont {S.~L.}\ \bibnamefont
  {Zhang}}, \bibinfo {author} {\bibfnamefont {G.}~\bibnamefont {van~der Laan}},
  \bibinfo {author} {\bibfnamefont {W.~W.}\ \bibnamefont {Wang}}, \bibinfo
  {author} {\bibfnamefont {A.~A.}\ \bibnamefont {Haghighirad}}, \ and\ \bibinfo
  {author} {\bibfnamefont {T.}~\bibnamefont {Hesjedal}},\ }\href {\doibase
  10.1103/PhysRevLett.120.227202} {\bibfield  {journal} {\bibinfo  {journal}
  {Phys. Rev. Lett.}\ }\textbf {\bibinfo {volume} {120}},\ \bibinfo {pages}
  {227202} (\bibinfo {year} {2018})}\BibitemShut {NoStop}%
\bibitem [{\citenamefont {Legrand}\ \emph {et~al.}(2018)\citenamefont {Legrand}
  \emph {et~al.}}]{Legrand_2018}%
  \BibitemOpen
  \bibfield  {author} {\bibinfo {author} {\bibfnamefont {W.}~\bibnamefont
  {Legrand}} \emph {et~al.},\ }\href {\doibase 10.1126/sciadv.aat0415}
  {\bibfield  {journal} {\bibinfo  {journal} {Science Advances}\ }\textbf
  {\bibinfo {volume} {4}} (\bibinfo {year} {2018}),\
  10.1126/sciadv.aat0415}\BibitemShut {NoStop}%
\bibitem [{\citenamefont {Zhang}\ \emph {et~al.}(2020)\citenamefont {Zhang}
  \emph {et~al.}}]{Zhang2020}%
  \BibitemOpen
  \bibfield  {author} {\bibinfo {author} {\bibfnamefont {S.}~\bibnamefont
  {Zhang}} \emph {et~al.},\ }\href {\doibase 10.1021/acs.nanolett.9b05141}
  {\bibfield  {journal} {\bibinfo  {journal} {Nano Letters}\ }\textbf {\bibinfo
  {volume} {20}},\ \bibinfo {pages} {1428} (\bibinfo {year}
  {2020})}\BibitemShut {NoStop}%
\bibitem [{\citenamefont {Seki}\ \emph {et~al.}(2012)\citenamefont {Seki},
  \citenamefont {Yu}, \citenamefont {Ishiwata},\ and\ \citenamefont
  {Tokura}}]{Seki2012}%
  \BibitemOpen
  \bibfield  {author} {\bibinfo {author} {\bibfnamefont {S.}~\bibnamefont
  {Seki}}, \bibinfo {author} {\bibfnamefont {X.~Z.}\ \bibnamefont {Yu}},
  \bibinfo {author} {\bibfnamefont {S.}~\bibnamefont {Ishiwata}}, \ and\
  \bibinfo {author} {\bibfnamefont {Y.}~\bibnamefont {Tokura}},\ }\href
  {\doibase 10.1126/science.1214143} {\bibfield  {journal} {\bibinfo  {journal}
  {Science}\ }\textbf {\bibinfo {volume} {336}},\ \bibinfo {pages} {198}
  (\bibinfo {year} {2012})}\BibitemShut {NoStop}%
\bibitem [{\citenamefont {Vlietstra}\ \emph {et~al.}(2013)\citenamefont
  {Vlietstra}, \citenamefont {Shan}, \citenamefont {Castel}, \citenamefont {van
  Wees},\ and\ \citenamefont {Ben~Youssef}}]{Vlietstra2013}%
  \BibitemOpen
  \bibfield  {author} {\bibinfo {author} {\bibfnamefont {N.}~\bibnamefont
  {Vlietstra}}, \bibinfo {author} {\bibfnamefont {J.}~\bibnamefont {Shan}},
  \bibinfo {author} {\bibfnamefont {V.}~\bibnamefont {Castel}}, \bibinfo
  {author} {\bibfnamefont {B.~J.}\ \bibnamefont {van Wees}}, \ and\ \bibinfo
  {author} {\bibfnamefont {J.}~\bibnamefont {Ben~Youssef}},\ }\href {\doibase
  10.1103/PhysRevB.87.184421} {\bibfield  {journal} {\bibinfo  {journal} {Phys.
  Rev. B}\ }\textbf {\bibinfo {volume} {87}},\ \bibinfo {pages} {184421}
  (\bibinfo {year} {2013})}\BibitemShut {NoStop}%
\bibitem [{\citenamefont {Nakayama}\ \emph {et~al.}(2013)\citenamefont
  {Nakayama} \emph {et~al.}}]{Nakayama2013}%
  \BibitemOpen
  \bibfield  {author} {\bibinfo {author} {\bibfnamefont {H.}~\bibnamefont
  {Nakayama}} \emph {et~al.},\ }\href {\doibase 10.1103/PhysRevLett.110.206601}
  {\bibfield  {journal} {\bibinfo  {journal} {Phys. Rev. Lett.}\ }\textbf
  {\bibinfo {volume} {110}},\ \bibinfo {pages} {206601} (\bibinfo {year}
  {2013})}\BibitemShut {NoStop}%
\bibitem [{\citenamefont {Althammer}\ \emph {et~al.}(2013)\citenamefont
  {Althammer} \emph {et~al.}}]{Althammer2013}%
  \BibitemOpen
  \bibfield  {author} {\bibinfo {author} {\bibfnamefont {M.}~\bibnamefont
  {Althammer}} \emph {et~al.},\ }\href {\doibase 10.1103/PhysRevB.87.224401}
  {\bibfield  {journal} {\bibinfo  {journal} {Phys. Rev. B}\ }\textbf {\bibinfo
  {volume} {87}},\ \bibinfo {pages} {224401} (\bibinfo {year}
  {2013})}\BibitemShut {NoStop}%
\bibitem [{\citenamefont {Aqeel}\ \emph {et~al.}(2015)\citenamefont {Aqeel}
  \emph {et~al.}}]{Aqeel2015}%
  \BibitemOpen
  \bibfield  {author} {\bibinfo {author} {\bibfnamefont {A.}~\bibnamefont
  {Aqeel}} \emph {et~al.},\ }\href {\doibase 10.1103/PhysRevB.92.224410}
  {\bibfield  {journal} {\bibinfo  {journal} {Phys. Rev. B}\ }\textbf {\bibinfo
  {volume} {92}},\ \bibinfo {pages} {224410} (\bibinfo {year}
  {2015})}\BibitemShut {NoStop}%
\bibitem [{\citenamefont {Ganzhorn}\ \emph {et~al.}(2016)\citenamefont
  {Ganzhorn} \emph {et~al.}}]{Ganzhorn2016}%
  \BibitemOpen
  \bibfield  {author} {\bibinfo {author} {\bibfnamefont {K.}~\bibnamefont
  {Ganzhorn}} \emph {et~al.},\ }\href {\doibase 10.1103/PhysRevB.94.094401}
  {\bibfield  {journal} {\bibinfo  {journal} {Phys. Rev. B}\ }\textbf {\bibinfo
  {volume} {94}},\ \bibinfo {pages} {094401} (\bibinfo {year}
  {2016})}\BibitemShut {NoStop}%
\bibitem [{\citenamefont {Aqeel}\ \emph {et~al.}(2016)\citenamefont {Aqeel}
  \emph {et~al.}}]{Aqeel2016}%
  \BibitemOpen
  \bibfield  {author} {\bibinfo {author} {\bibfnamefont {A.}~\bibnamefont
  {Aqeel}} \emph {et~al.},\ }\href {\doibase 10.1103/PhysRevB.94.134418}
  {\bibfield  {journal} {\bibinfo  {journal} {Phys. Rev. B}\ }\textbf {\bibinfo
  {volume} {94}},\ \bibinfo {pages} {134418} (\bibinfo {year}
  {2016})}\BibitemShut {NoStop}%
\bibitem [{\citenamefont {Aqeel}\ \emph {et~al.}(2017)\citenamefont {Aqeel},
  \citenamefont {Mostovoy}, \citenamefont {van Wees},\ and\ \citenamefont
  {Palstra}}]{Aqeel2017}%
  \BibitemOpen
  \bibfield  {author} {\bibinfo {author} {\bibfnamefont {A.}~\bibnamefont
  {Aqeel}}, \bibinfo {author} {\bibfnamefont {M.}~\bibnamefont {Mostovoy}},
  \bibinfo {author} {\bibfnamefont {B.~J.}\ \bibnamefont {van Wees}}, \ and\
  \bibinfo {author} {\bibfnamefont {T.~T.~M.}\ \bibnamefont {Palstra}},\ }\href
  {http://stacks.iop.org/0022-3727/50/i=17/a=174006} {\bibfield  {journal}
  {\bibinfo  {journal} {Journal of Physics D: Applied Physics}\ }\textbf
  {\bibinfo {volume} {50}},\ \bibinfo {pages} {174006} (\bibinfo {year}
  {2017})}\BibitemShut {NoStop}%
\bibitem [{\citenamefont {Ji}\ \emph {et~al.}(2017)\citenamefont {Ji} \emph
  {et~al.}}]{Ji2017}%
  \BibitemOpen
  \bibfield  {author} {\bibinfo {author} {\bibfnamefont {Y.}~\bibnamefont {Ji}}
  \emph {et~al.},\ }\href {\doibase 10.1063/1.4989680} {\bibfield  {journal}
  {\bibinfo  {journal} {Applied Physics Letters}\ }\textbf {\bibinfo {volume}
  {110}},\ \bibinfo {pages} {262401} (\bibinfo {year} {2017})}\BibitemShut
  {NoStop}%
\bibitem [{\citenamefont {Hoogeboom}\ \emph {et~al.}(2017)\citenamefont
  {Hoogeboom}, \citenamefont {Aqeel}, \citenamefont {Kuschel}, \citenamefont
  {Palstra},\ and\ \citenamefont {van Wees}}]{Hoogeboom2017}%
  \BibitemOpen
  \bibfield  {author} {\bibinfo {author} {\bibfnamefont {G.~R.}\ \bibnamefont
  {Hoogeboom}}, \bibinfo {author} {\bibfnamefont {A.}~\bibnamefont {Aqeel}},
  \bibinfo {author} {\bibfnamefont {T.}~\bibnamefont {Kuschel}}, \bibinfo
  {author} {\bibfnamefont {T.~T.~M.}\ \bibnamefont {Palstra}}, \ and\ \bibinfo
  {author} {\bibfnamefont {B.~J.}\ \bibnamefont {van Wees}},\ }\href {\doibase
  10.1063/1.4997588} {\bibfield  {journal} {\bibinfo  {journal} {Applied
  Physics Letters}\ }\textbf {\bibinfo {volume} {111}},\ \bibinfo {pages}
  {052409} (\bibinfo {year} {2017})}\BibitemShut {NoStop}%
\bibitem [{\citenamefont {Wang}\ \emph {et~al.}(2017)\citenamefont {Wang} \emph
  {et~al.}}]{Wang2017}%
  \BibitemOpen
  \bibfield  {author} {\bibinfo {author} {\bibfnamefont {H.}~\bibnamefont
  {Wang}} \emph {et~al.},\ }\href {\doibase 10.1063/1.4986372} {\bibfield
  {journal} {\bibinfo  {journal} {Journal of Applied Physics}\ }\textbf
  {\bibinfo {volume} {122}},\ \bibinfo {pages} {083907} (\bibinfo {year}
  {2017})}\BibitemShut {NoStop}%
\bibitem [{\citenamefont {Fischer}\ \emph {et~al.}(2018)\citenamefont {Fischer}
  \emph {et~al.}}]{Fischer2018}%
  \BibitemOpen
  \bibfield  {author} {\bibinfo {author} {\bibfnamefont {J.}~\bibnamefont
  {Fischer}} \emph {et~al.},\ }\href {\doibase 10.1103/PhysRevB.97.014417}
  {\bibfield  {journal} {\bibinfo  {journal} {Phys. Rev. B}\ }\textbf {\bibinfo
  {volume} {97}},\ \bibinfo {pages} {014417} (\bibinfo {year}
  {2018})}\BibitemShut {NoStop}%
\bibitem [{\citenamefont {Lebrun}\ \emph {et~al.}(2019)\citenamefont {Lebrun}
  \emph {et~al.}}]{Lebrun2019}%
  \BibitemOpen
  \bibfield  {author} {\bibinfo {author} {\bibfnamefont {R.}~\bibnamefont
  {Lebrun}} \emph {et~al.},\ }\href {\doibase 10.1038/s42005-019-0150-8}
  {\bibfield  {journal} {\bibinfo  {journal} {Communications Physics}\ }\textbf
  {\bibinfo {volume} {2}},\ \bibinfo {pages} {50} (\bibinfo {year}
  {2019})}\BibitemShut {NoStop}%
\bibitem [{\citenamefont {Jia}\ \emph {et~al.}(2011)\citenamefont {Jia},
  \citenamefont {Liu}, \citenamefont {Xia},\ and\ \citenamefont
  {Bauer}}]{JiaSTT2011}%
  \BibitemOpen
  \bibfield  {author} {\bibinfo {author} {\bibfnamefont {X.}~\bibnamefont
  {Jia}}, \bibinfo {author} {\bibfnamefont {K.}~\bibnamefont {Liu}}, \bibinfo
  {author} {\bibfnamefont {K.}~\bibnamefont {Xia}}, \ and\ \bibinfo {author}
  {\bibfnamefont {G.~E.~W.}\ \bibnamefont {Bauer}},\ }\href
  {http://stacks.iop.org/0295-5075/96/i=1/a=17005?key=crossref.b410de24aa763d1d65ff0d713a3374c9}
  {\bibfield  {journal} {\bibinfo  {journal} {Europhys. Lett.}\ }\textbf
  {\bibinfo {volume} {96}},\ \bibinfo {pages} {17005} (\bibinfo {year}
  {2011})}\BibitemShut {NoStop}%
\bibitem [{\citenamefont {Chen}\ \emph {et~al.}(2013)\citenamefont {Chen} \emph
  {et~al.}}]{Chen2013}%
  \BibitemOpen
  \bibfield  {author} {\bibinfo {author} {\bibfnamefont {Y.-T.}\ \bibnamefont
  {Chen}} \emph {et~al.},\ }\href {\doibase 10.1103/PhysRevB.87.144411}
  {\bibfield  {journal} {\bibinfo  {journal} {Phys. Rev. B}\ }\textbf {\bibinfo
  {volume} {87}},\ \bibinfo {pages} {144411} (\bibinfo {year}
  {2013})}\BibitemShut {NoStop}%
\bibitem [{\citenamefont {Vlietstra}\ \emph {et~al.}(2014)\citenamefont
  {Vlietstra} \emph {et~al.}}]{Vlietstra2014}%
  \BibitemOpen
  \bibfield  {author} {\bibinfo {author} {\bibfnamefont {N.}~\bibnamefont
  {Vlietstra}} \emph {et~al.},\ }\href {\doibase 10.1103/PhysRevB.90.174436}
  {\bibfield  {journal} {\bibinfo  {journal} {Phys. Rev. B}\ }\textbf {\bibinfo
  {volume} {90}},\ \bibinfo {pages} {174436} (\bibinfo {year}
  {2014})}\BibitemShut {NoStop}%
\bibitem [{\citenamefont {Schwarze}\ \emph {et~al.}(2015)\citenamefont
  {Schwarze} \emph {et~al.}}]{Schwarze_2015}%
  \BibitemOpen
  \bibfield  {author} {\bibinfo {author} {\bibfnamefont {T.}~\bibnamefont
  {Schwarze}} \emph {et~al.},\ }\href {\doibase 10.1038/nmat4223} {\bibfield
  {journal} {\bibinfo  {journal} {Nat. Mat.}\ }\textbf {\bibinfo {volume}
  {14}},\ \bibinfo {pages} {478–483} (\bibinfo {year} {2015})}\BibitemShut
  {NoStop}%
\bibitem [{\citenamefont {Hanke}\ \emph {et~al.}(2020)\citenamefont {Hanke}
  \emph {et~al.}}]{Hanke2020}%
  \BibitemOpen
  \bibfield  {author} {\bibinfo {author} {\bibfnamefont {J.-P.}\ \bibnamefont
  {Hanke}} \emph {et~al.},\ }\href {\doibase 10.1103/PhysRevB.101.014428}
  {\bibfield  {journal} {\bibinfo  {journal} {Phys. Rev. B}\ }\textbf {\bibinfo
  {volume} {101}},\ \bibinfo {pages} {014428} (\bibinfo {year}
  {2020})}\BibitemShut {NoStop}%
\bibitem [{\citenamefont {Keesman}\ \emph {et~al.}(2015)\citenamefont {Keesman}
  \emph {et~al.}}]{Keesman_2015}%
  \BibitemOpen
  \bibfield  {author} {\bibinfo {author} {\bibfnamefont {R.}~\bibnamefont
  {Keesman}} \emph {et~al.},\ }\href {\doibase 10.1103/PhysRevB.92.134405}
  {\bibfield  {journal} {\bibinfo  {journal} {Phys. Rev. B}\ }\textbf {\bibinfo
  {volume} {92}},\ \bibinfo {pages} {134405} (\bibinfo {year}
  {2015})}\BibitemShut {NoStop}%
\end{thebibliography}

%

\end{document}